\documentclass{raa}            % referee version: for submission
\usepackage{graphicx,times}             %for PS/EPS graphics inclusion, new
\usepackage{threeparttable}
\usepackage[rotateright]{rotating}
\usepackage{lscape}

\input{epsf.sty}                        %for PS/EPS graphics inclusion, old
\input{psfig.sty}                       %for PS/EPS graphics inclusion, old

\begin{document}

\title{Disappearance of a coronal hole induced by a filament activation}
%   \subtitle{I. Place Your Subtitle Here}

   \volnopage{Vol.0 (200x) No.0, 000--000}
   \setcounter{page}{1}

\author{Lin Ma\inst{1,2}
	\and Zhong-Quan Qu\inst{1}
	\and Xiao-Li Yan\inst{1}
	\and Zhi-Ke Xue\inst{1}}

\institute{Yunnan Observatories, Chinese Academy of Sciences, Kunming 650011, China; {\itshape{malin-567@ynao.ac.cn}}\\
 \and University of Chinese Academy of Sciences, Beijing 100049, China}

 \date{Received~~ month day; accepted~~ ~~month day}

\abstract{we present a rare observation of direct magnetic interaction between an activating filament and a coronal hole (CH).
The filament was a quiescent one located at the northwest of the CH. It underwent a nonradial activation, during which filament material constantly fell and intruded into the CH. As a result, the CH was clearly destroyed by the intrusion. Brightenings appeared at the boundaries and in the interior of the CH, meanwhile, its west boundaries began to retreat and the area gradually shrank. It is noted that the CH went on shrinking after the end of the intrusion and finally disappeared entirely. Following the filament activation, three coronal dimmings (D1-D3) were formed, among which D1 and D2 persisted throughout the complete disappearance of the CH. The derived coronal magnetic configuration shows that the filament was located below an extended loop system which obviously linked D1 to D2. By comparison with this result of extrapolation, our observations imply that the interaction between the filament and the CH involved direct intrusion of the filament material to the CH and the disappearance of the CH might be due to interchange reconnection between the expanding loop system and the CH's open field.
}
\authorrunning{Lin Ma et al}
\titlerunning{Disappearance of a coronal hole induced by a filament activation}

\maketitle

\section{INTRODUCTION}
\label{sect:intro}
Coronal holes (CHs) are regions of magnetic fields dominated by one polarity and the magnetic lines open outwards to
interplanetary space. Plasma in CHs is evacuated into interplanetary space along the open field lines, giving rise to high-speed solar winds which may induce geomagnetic storms on Earth (Krieger et al. 1973; Fisk \& Schwadron 2001; Tu et al. 2005). Owing to reduced density and temperature, CHs appear as dark areas when observed with X-ray and extreme-ultraviolet (EUV) lines (Munro \& Withbroe 1972; Vaiana et al. 1976). According to their locations and lifetimes, CHs can be grouped into three categories: polar, isolated, and transient. At the minimum of the solar activity cycle, the solar atmosphere is dominated by two polar CHs. They can exist for several years. The isolated CHs have an occurrence rate that follows the solar activity cycle and can survive for many solar rotations. The transient ones are often associated with eruption events, persisting for only several days. The CH boundaries (CHBs) separate two topologically different magnetic field configurations: open field in CHs and closed field in active regions (ARs) or the quiet Sun. It has been suggested that magnetic reconnection must occur at CHBs to maintain the CH integrity (Wang \& Sheeley 1994; Fisk et al. 1999; Kahler \& Hudson 2002).

Recently, such reconnection at CHBs has generally been accepted as ``interchange reconnection'' (Crooker et al. 2002), of which the result is an exchange of footpoints between open and closed field lines, with conservation of the total open or closed flux. Wang \& sheeley (2004) presented two kinds of interchange reconnection, respectively called X-type and Y-type interchange reconnections in this paper. The X-type reconnection typically occurs when a magnetic bipole emerges inside or at the boundary of a CH. It transfers the magnetic flux through an X-point which marks the intersection between the open flux and the closed flux. The Y-type interchange reconnection is inherently three-dimensional, involving interaction between an open field line and a neighboring but non-coplanar closed loop. It occurs at the apex of the closed loop and leads to stepwise displacements within a region of single magnetic polarity. Hitherto, observational evidence of interchange reconnection is rather rare. Madjarska et al. (2004) provided for the first time observational evidence for interchange reconnection, which was bidirectional jets occurring along CHBs. Baker et al. (2007) studied the interchange reconnection between a CH and an emerging AR. The reconnection was evidenced by the bright loops linking the CH to the AR, the retreat of the CHBs and the coronal dimming on one side of the AR. Then more direct evidence was given by Ma et al. (2014). They deemed that the bright loops forming between the CH and the AR, the boundary retreat of the CH, the coronal dimming on the negative polarity of the AR, the disturbance of the AR and the brightening and EUV jets appearing at the CHBs jointly represented credible evidence of interchange reconnection.

Coronal dimmings are a frequently occurring phenomenon in many eruption events. They appear as darkenings in EUV and brightenings in He $\textrm{I}$ 10830 {\AA} in regions of unipolar magnetic field, with lifetime of only a few hours to days, therefore, they are commonly referred to as ``transient coronal holes'' (Rust 1983; de Toma et al. 2005). Since typical timescales of the coronal dimmings are shorter than that of radiation in the corona, they are often explained as due to a density drop rather than a temperature decrease (Hudson et al. 1996). As mentioned by many researchers, coronal dimmings are caused by the expansion of coronal magnetic fields and thus mark the footprints of the expanded fields. It is increasingly clear that the formation of dimmings may represent the large-scale rearrangement of coronal magnetic fields and the dimmings can be regarded as key low-corona proxies of coronal mass ejections (CMEs) (Thompson et al. 2000; Jiang et al. 2011). Coronal dimmings can take complicated patterns in a wide range of scales. A pair of compact and symmetric dimmings with opposite magnetic polarities can form adjacent to the two ends of an eruptive filament and are believed to be the evacuated footprints of a large-scale flux rope ejection (Sterling \& Hudson 1997; Jiang et al. 2006). Dimmings associated with brightenings can form away from the eruptive source region and are considered to be signatures of the expansion of overlying magnetic fields (Manoharan et al. 1996).

In the magnetically dominant corona, a magnetic structure possibly interacts with other magnetic structures and different kinds of magnetic interactions could result in a broad form of solar activities. For instance, some works have suggested that interchange reconnection between CHs and ARs can dominate the large-scale evolution of CHBs (Wang \& Sheeley 2004; Baker et al. 2007; Ma et al. 2014); interaction of an erupting flux rope with overlying loops can produce remote brightenings and dimmings (Manoharan et al. 1996; Wang 2005); an EIT wave may interact with the ambient ARs or CHs and produces visible deflections (Thompson et al. 1999; Veronig et al. 2006); direct collision of an erupting filament with a CH can lead to change of the CH and reflection of the filament (Jiang et al. 2007); and a filament hit by a jet may erupt obliquely (Jiang et al. 2009). Therefore, observations of these magnetic interactions are favorable for understanding the complicated solar activity.

On 2010 November 13, a filament activation was observed by the Solar Dynamics Observatory (SDO) and the Global Oscillation Network Group (GONG; Harvey et al. 1996). During the activation, filament material constantly fell and intruded the CH. The CH then changed and finally disappeared. In this paper, we will present the interaction between the CH and the filament in detail.

\section{DATA ANALYSIS}

In the present study, we used full-disk EUV images with a pixel size of 0.$^{''}$6 from the Atmospheric Imaging Assembly (AIA; Lemen et al. 2012) and full-disk line-of-sight magnetograms with 45 s time cadence and a pixel size of 0.$^{''}$5 from the Helioseismic and Magnetic Imager (HMI; Schou et al. 2012). Both the AIA and the HMI are on board the SDO. The AIA employs two UV bands, one visible-light bands and seven narrow EUV bands centered on Fe $\textrm{XVIII}$ 94 {\AA}, Fe $\textrm{VIII}$, $\textrm{XXI}$ 131 {\AA}, Fe $\textrm{IX}$ 171 {\AA}, Fe $\textrm{XII}$, $\textrm{XXIV}$ 193 {\AA}, Fe $\textrm{XIV}$ 211 {\AA}, He $\textrm{II}$ 304 {\AA} and Fe $\textrm{XVI}$ 335 {\AA}. The HMI observes the full Sun at six polarization states in the Fe $\textrm{I}$ 6173 {\AA} absorption line. For this event, we adopted AIA 304 {\AA} and 193 {\AA} data at Level 1.5 with 12 s cadence. Moreover, the full-disk H$\rm{\alpha}$ images from the GONG were included to identify the filament. To inspect whether or not flares or CMEs took place in the event, we used the soft X-ray (SXR) light curves observed by the Geostationary Operational Environmental Satellite (GOES), as well as the C2 and C3 white-light coronagraph data from the Large Angle and Spectrometric Coronagraph (LASCO; Brueckner et al. 1995).

Finally, all of the images were differentially rotated to a reference time (00:00 UT on 2010 November 13). The CHBs are defined as the regions with intensities 1.5 times the average intensity of the darkest region inside the CH (Madjarska \& Wiegelmann 2009; Yang et al. 2011).

%%%%%%%%%%%%%%%%%%%%%%%%%%%%%%%%%%%%%%%%%%%%%%%%%%%%%%%%%%%%%%%%%%%
\begin{figure}
\centering
\includegraphics[width=0.7\textwidth]{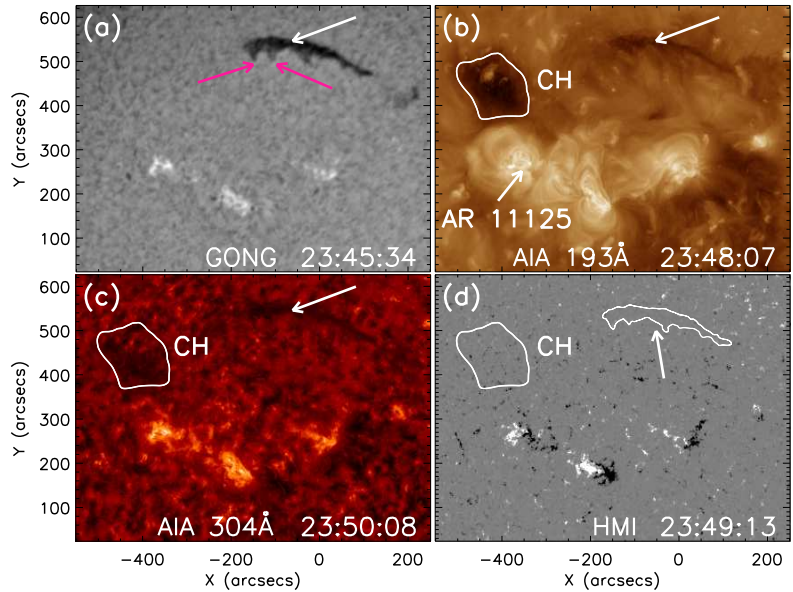}
\caption{GONG H$\rm{\alpha}$ (a), AIA 193 {\AA} (b) and AIA 304 {\AA} (c) images, and HMI magnetogram (d) with contours of the CH obtained from panel (b) and the filament obtained from panel (a). The white arrows indicate the filament. The magenta arrows in panel (a) indicate the two barbs of the filament. The field of view (FOV) is 800$^{''}$ $\times$ 600$^{''}$. }
\label{f1}
\end{figure}
									
\section{RESULTS}	

Figure 1 shows the general appearance of the filament and the CH before the filament activation. The white arrows in panels (a-d) indicate the filament. The white contours in panels (b-d) denote the CHBs. The filament was located on a quiet-sun region in the northern hemisphere, centering at N36$^{\circ}$E3$^{\circ}$. It was made up of spine and barbs, and two of the barbs (indicated by the magenta arrows in panel (a)) were observed as promoters of the activation. To the southeast of the filament, the CH was centered at about N29$^{\circ}$E28$^{\circ}$, with longitudinal extension of about 12$^{\circ}$ and latitudinal width of nearly 10$^{\circ}$. It was immediately adjacent to the northern boundary of AR 11125. When we superimposed the contours of the CH and the filament on the corresponding HMI magnetogram in Figure 1(d), it is clear that the CH was dominated by negative polarity and the filament lay along the polarity inversion line (PIL).

The filament began to activate at about 00:00 UT November 13. Figure 2 presents the detailed process of the filament activation and associated phenomena in GONG H$\rm{\alpha}$, AIA 304 {\AA} and AIA 193 {\AA} images. Our H$\rm{\alpha}$ observations did not completely cover the activation, but 304 {\AA} images showed an intact process which can be broken down into the following consecutive stages. (1) From 00:00 UT to 05:03 UT, the filament horizontally moved eastward. The filament activation started from a upward roll of the filament material in the two barbs (indicated by the magenta arrows in Figure 1(a)). The roll drove the filament to move eastward and thus made the two barbs fade away (panels (a1-a2) and (b1)). (2) Next, accompanying the eastward motion, filament material constantly fell from the east end of the filament. The falling filament material formed a curved structure which first dropped in on the west boundaries of the CH at about 05:06 UT and then expanded and moved eastward, gradually invading the CH (panels (b2-b4)). The white contours in panel (b2) indicate the CHBs. When we used the sequential labels, ``p1,'' ``p2'' and ``p3,'' to mark the position of the curved structure respectively at 05:06, 07:10 and 09:00 UT, the eastward shift of the curved structure can be easily found out in the 304 {\AA} images (see panels (b2-b4)). (3) From 09:00 UT to 10:10 UT, the filament branched out into two parts (indicated by the thin white arrows in panel (b4)): filament material in the east part went on falling and maintained the intrusion behaviour, while that in the west part retreated back. By 10:10 UT, the east part disappeared, implying that the intrusion of the filament to the CH ceased. Note that the filament became completely invisible at about 09:00 UT in the H$\rm{\alpha}$ image (panel (a3)). (4) The remaining west part then wore off and eventually disappeared at about 14:12 UT. Because the filament did not erupt in our observations, its disappearance should be thermo-disappearance.

In AIA 193 {\AA} observations, the filament activation showed two remarkable characteristics (see panels (c1-c3)). One is that the fall and intrusion of the filament were accompanied by a patch of EUV brightenings (indicated by the thick white arrows in panels (c1-c2)). The brightenings stood along the falling curved structure, initially appearing at the west boundaries of the CH and progressively extending into the CH interior, and finally disappeared after the end of the intrusion (10:10 UT). The other is that three coronal dimming regions were formed during the filament activation. By comparison with HMI magnetogram (panel (d) in Figure 1), we can see that they are located on opposite-polarity sides of the filament: a negative-polarity region, ``D1,'' and two positive-polarity ones, ``D2'' and ``D3''. Our observations suggest that some interaction occurred between the activating filament and the CH. It was due to the obvious intrusion of the filament to the CH and the timely appearance of the brightenings that we can pay attention to such interaction.
%%%%%%%%%%%%%%%%%%%%%%%%%%%%%%%%%%%%%%%%%%%%%%%%%%%%%%%%%%%%%%%%%%%%%%%%%%%%
\begin{figure}
\centering
\includegraphics[width=\textwidth]{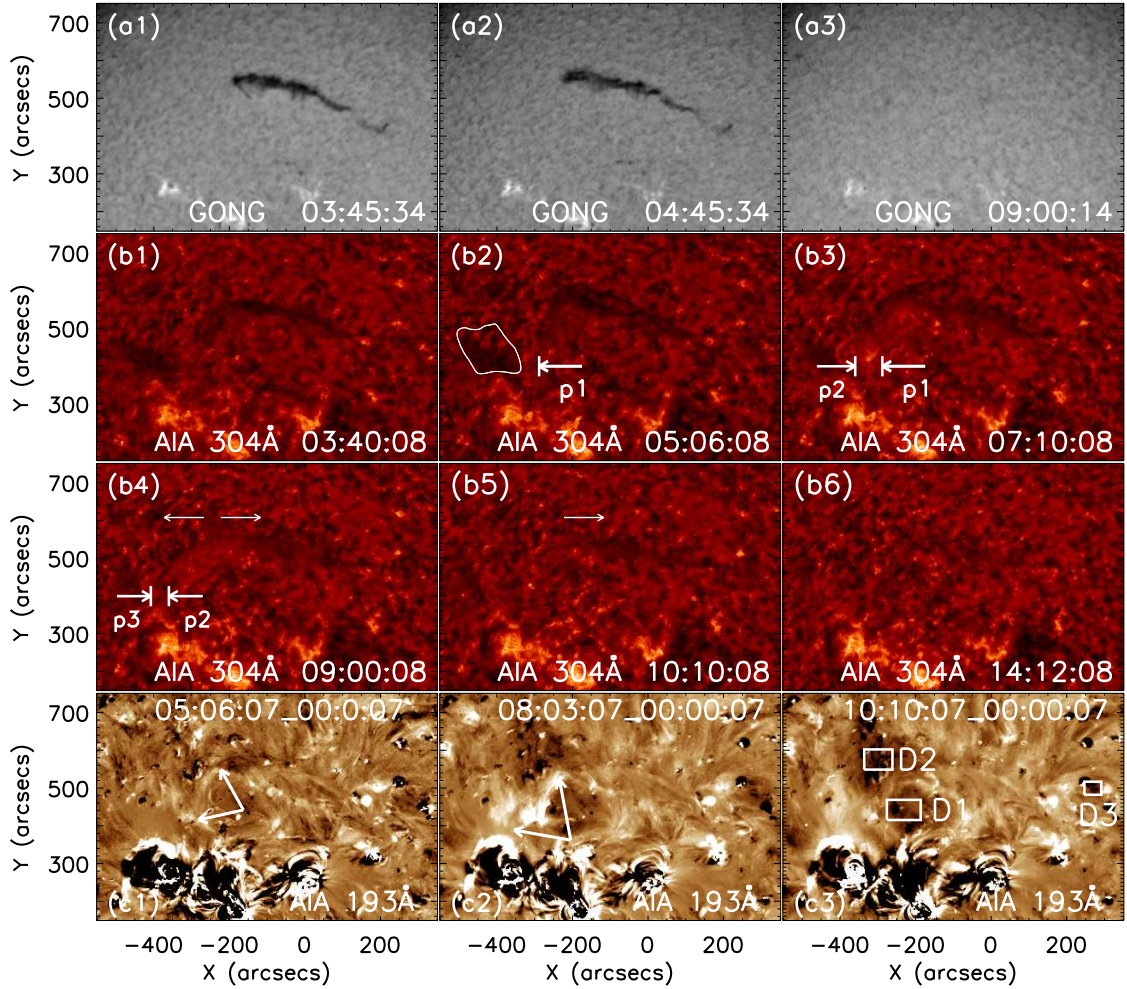}
\caption{GONG H$\rm{\alpha}$ (a1-a3), AIA 304 {\AA} (b1-b6) and AIA 193 {\AA} fixed-base difference (c1-c3) images. The white contour in panel (b2) indicates the CHBs obtained from the 193 {\AA} image at 05:06 UT. The white arrows, ``p1,'' ``p2'' and ``p3'' show the position of the curved structure. The thin white arrows in panels (b4-b5) indicate the directions of the two parts of the filament. Brightenings (white arrows in panel (c1-c2)) and three coronal dimmings, ``D1,'' ``D2'' and ``D3'' were formed during the filament activation. The FOV is 900$^{''}$ $\times$ 600$^{''}$.}
\label{f2}
\end{figure}
%%%%%%%%%%%%%%%%%%%%%%%%%%%%%%%%%%%%%%%%%%%%%%%%%%%%%%%%%%%%%%%%%%%%%%	

In order to study the interaction between the CH and the filament and the response of the CH to it in detail, Figure 3 presents the close-up view of CH's evolution in AIA 193 {\AA} images. In the figure, the CH appears as a dark region, of which the boundaries are delineated by white contours. Regarding 05:06 UT as the starting time of the filament intrusion, the CH was almost unchanged theretofore (see panels (a1-a2)). Then, following the fall and intrusion of the filament material, the CH was clearly destroyed (see panels (a2-a5)). About from 05:06 UT, EUV brightenings appeared at the boundaries and in the interior of the CH (white arrows in panels (a3-a4) and Figure 2(c1-c2)), meanwhile, the west boundaries of the CH constantly retreated and its area gradually shrank (see panels (a3-a5)). This possibly suggests that some adjustments occurred in the open magnetic field of the CH. At about 10:10 UT, the filament material stopped to fall and intrude, and major part of the CH and the brightenings disappeared (see panel (a5)). However, after that the retreat of the CHBs did not stop, and the remainder of the CH went on shrinking (see panels (a6-a7)) and ultimately disappeared at about 20:30 UT (panel (a8)). It appears that the interaction between the filament and the CH still persisted after the end of the intrusion and the filament disappearance, implying that it was not simply due to the filament intrusion and inevitably contained other process. Since D1 and D2 still can be discerned in the 193 {\AA} fixed-base difference image (panel (a9)) after the CH disappearance, it is probably that the interaction was in connection with the dimming formation process. We believe that such seasonable change of the CH represented a creditable manifestation of the interaction.

In Figure 4(a), a space-time plot along the slit ``s1-s2'' (shown as the white dashed line in Figure 3(a1)) is displayed to examine the CH's evolution. We obtained the profile in the plot by averaging five pixels in 193 {\AA} images in the direction perpendicular to the slit over time from 00:00 UT to 20:00 UT. Consistent with the observations, the plot shows apparent retreat of the CHBs and appearance of brightenings just after the fall of filament material. In Figure 4(b), the light curves of AIA 193 {\AA} intensities in the areas D1, D2 and D3 are plotted. It was found that the intensity drops in D1, D2 and D3 occurred at 04:03, 04:48 and 05:00 UT, respectively. This implies that the dimming formation was correlated with the filament activation. However, in this event, we did not find any significant GOES or H$\rm{\alpha}$ flares and LASCO CMEs corresponding to the filament activation.									

\begin{figure}
\centering
\includegraphics[width=14cm]{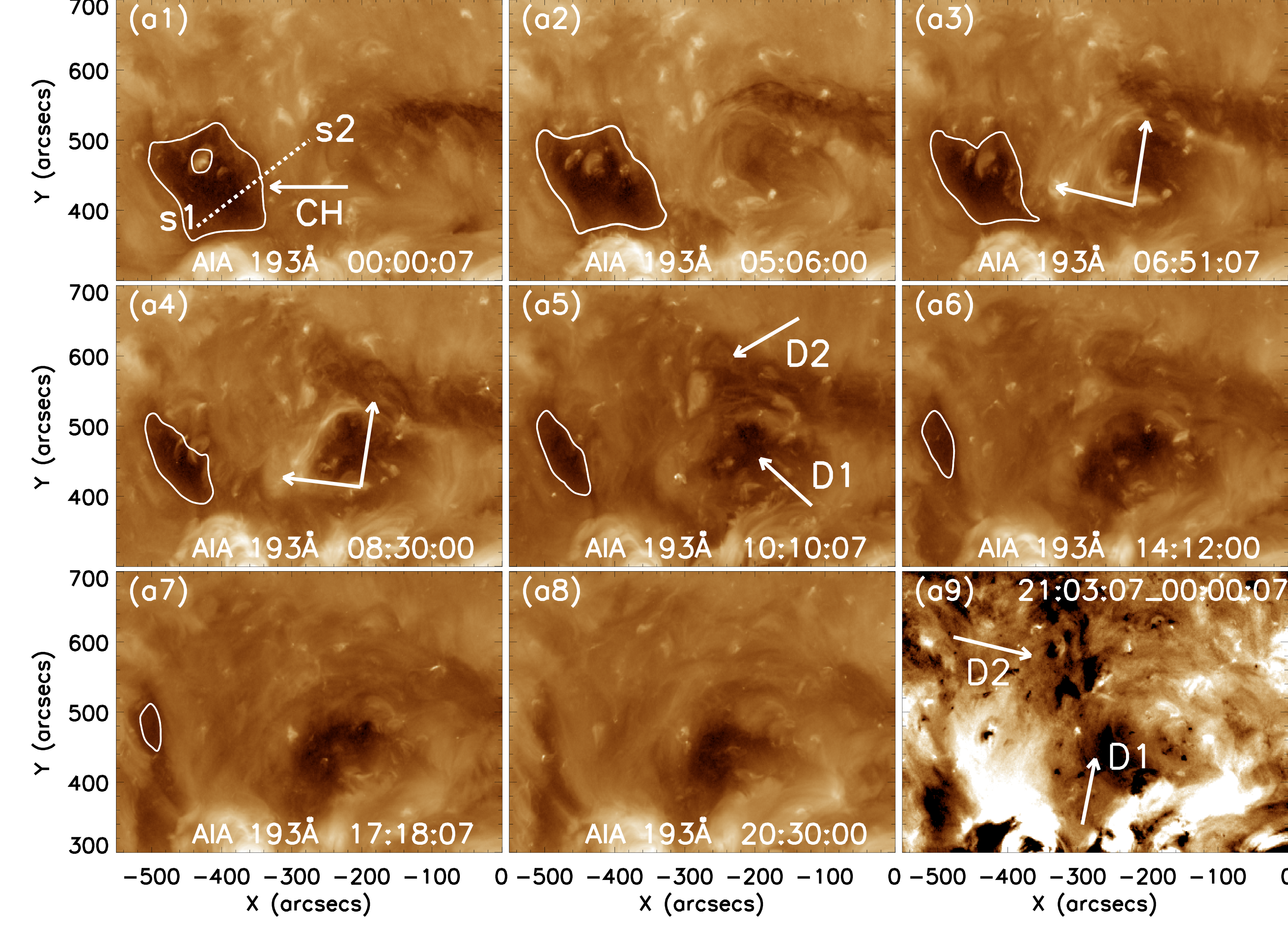}
\caption{AIA 193 {\AA} direct (a1-a8) and fixed-base difference (a9) images. The white contours represent CHBs.
The slit ``s1-s2'' (white dashed line in panel (a1)) denotes the position along which the space-time plot (shown in Figure 4(a)) is obtained. The white arrows in panels (a3-a4) indicate the brightenings. The FOV is 550$^{''}$ $\times$ 400$^{''}$.}
\label{f3}
\end{figure}
%%%%%%%%%%%%%%%%%%%%%%%%%%%%%%%%%%%%%%%%%%%%%%%%%%%%%%%%%%%%%%%%%%%%%%%%%
\begin{figure}
\centering
\includegraphics[width=6cm]{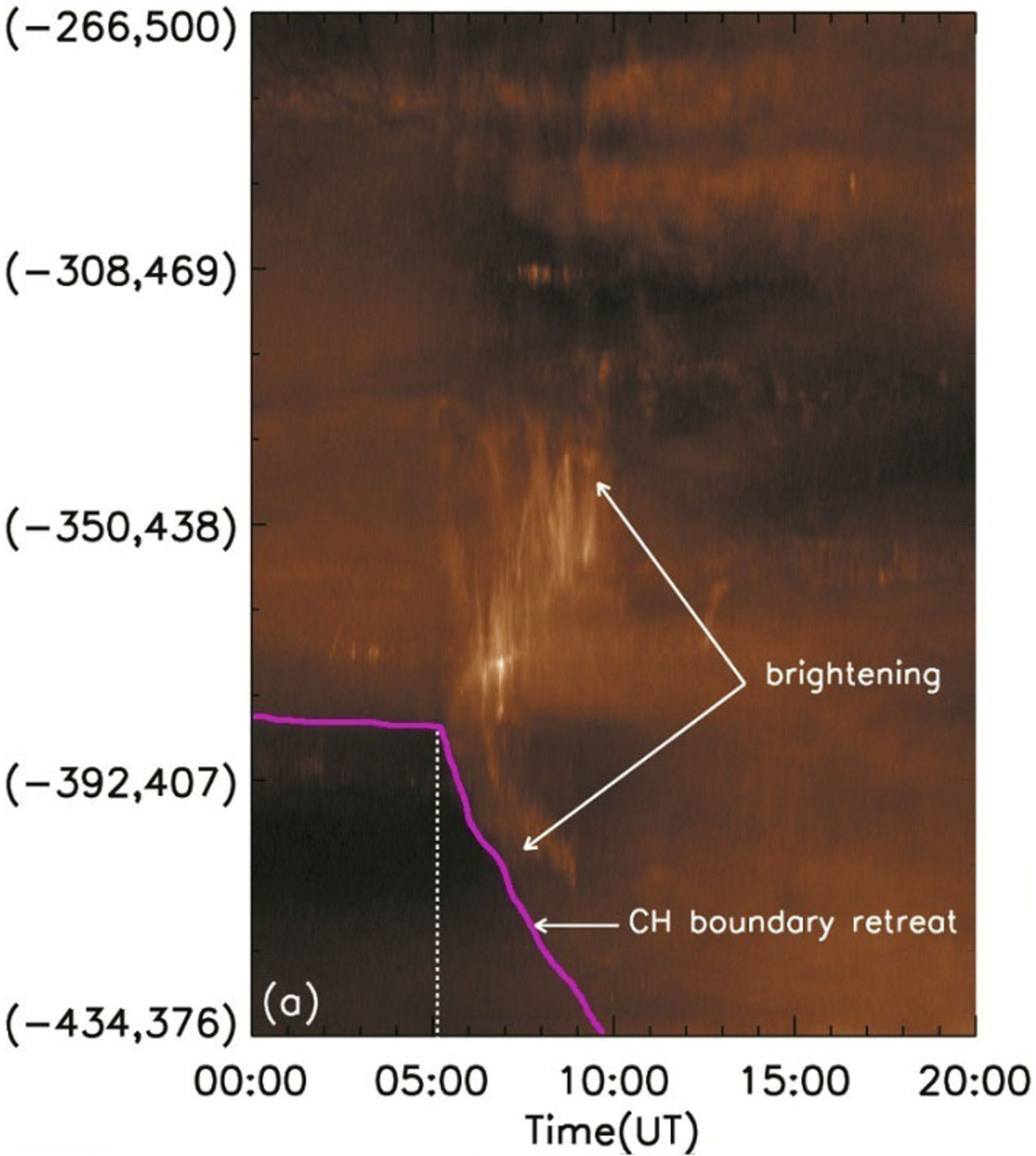}
\includegraphics[width=8.3cm]{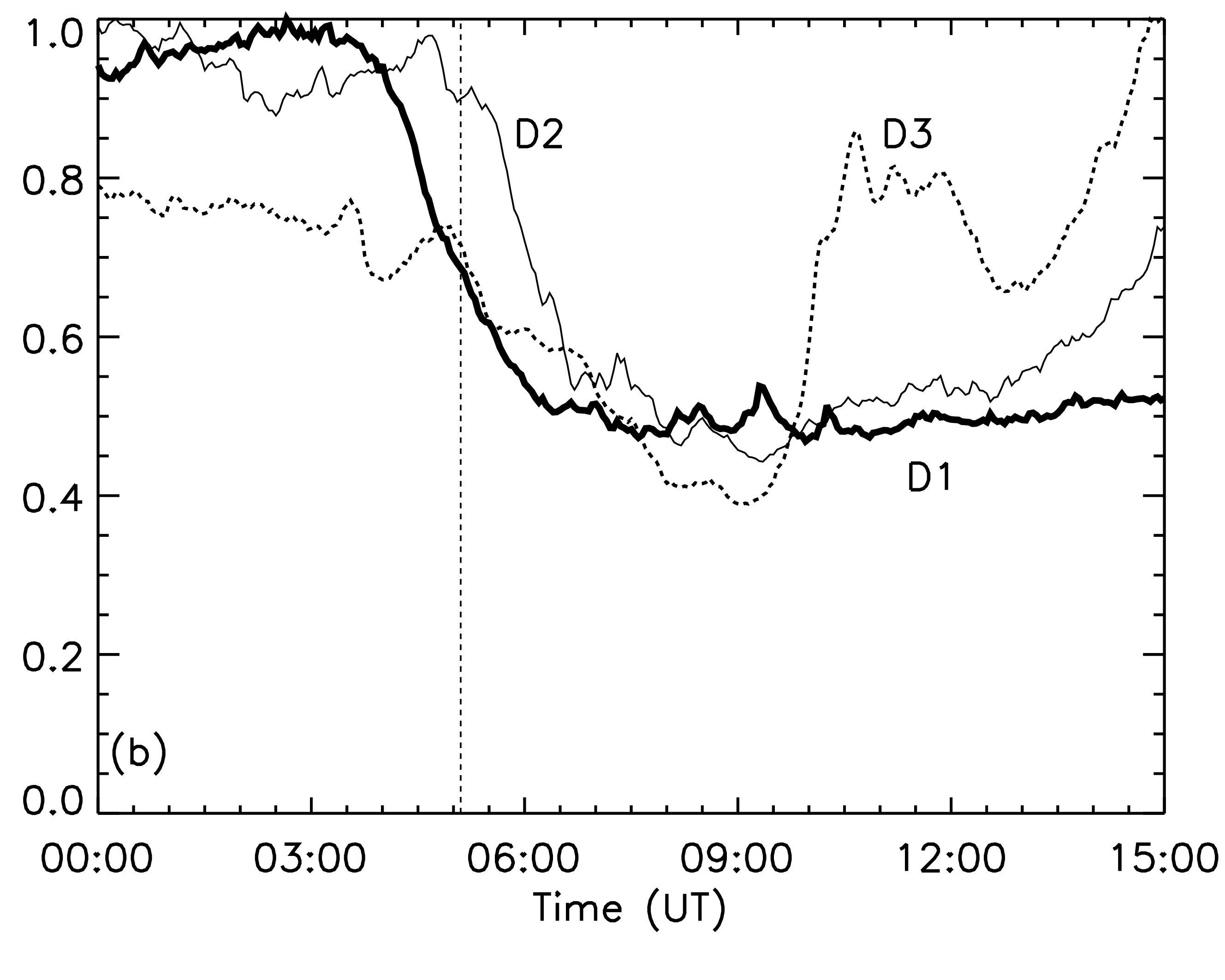}
\caption{Space-time plot (a) along the slit ``s1-s2'' (shown in Figure 3(a1)) and the light curves of AIA 193 {\AA} intensities in areas centered on D1 (the thick solid curve in (b)), D2 (the thin solid curve in (b)) and D3 (the dashed curve in (b)). The purple curve in panel (a) delineates CHBs. The light curves are computed from the intensity integrated and normalized over the regions indicated by the white boxes in Figure 2(c3). The vertical dashed lines indicate the start time of the fall of filament material.}
\label{f4}
\end{figure}
%%%%%%%%%%%%%%%%%%%%%%%%%%%%%%%%%%%%%%%%%%%%%%%%%%%%%%%%%%%%%%%%%%%%%%%%%

\section{MAGNETIC CONFIGURATION ANALYSIS}

According to the observations described above, we suspect the following scenario: the filament activation first led to the expansion of the overlying magnetic loop system, with footprints represented by D1 and D2. Then, accompanying the filament intrusion, the expanding loop system met and interacted with the CH's open field. Both the interaction and the filament intrusion could lead to the change of the CH. After the end of the filament intrusion, the interaction still persisted and finally resulted in the CH disappearance. To confirm our conjecture and understand the interaction between the CH and the filament more clearly, the magnetic connectivity of D1 and D2 and the knowledge of the large-scale coronal configuration are crucial. Using the results of the Potential-Field Source-Surface (PFSS) model (Schrijver \& DeRosa 2003) based on the synoptic magnetic maps from the Michelson Doppler Imager (MDI; Scherrer et al. 1995), we construct the coronal magnetic field and estimate its variation.
											
In Figure 5, the typical coronal magnetic field lines are presented. The outlines of the three dimmings (D1-D3) represented by the white boxes, and the PIL (white cross line) are also superposed. There are two kinds of magnetic field lines in the figure. The green lines stand for open field lines of the CH, while the yellow lines refer to closed field lines. In addition to those confined arcades holding the ARs, an extended loop system, labelled with ``L'', is distinguished in this configuration. L is overlying the filament and obviously linking D1 to D2. According to the magnetic connectivity and the observational results that the CH continued to shrink after the end of the filament intrusion and D1 and D2 persisted throughout its whole disappearance, we speculate that the interaction between the CH and the filament should contain two processes. First, direct intrusion of filament material to the CH would produce brightenings, just like waves invading into CHs. As a result, the CH changed seasonably to adjust the appearance of the brightenings. Second, simple expansion of L, pushed eastward by the filament activation, would produce D1 and D2 and drive interaction with the CH's open field. The interaction could close down the CH field, leading to the boundary retreat and area shrinkage of the CH. Both of the two processes took effect during the early phase of the interaction between the CH and the filament. However, after the end of the filament intrusion, only the interaction between the expanding L and the CH's open field persisted and finally resulted in the CH disappearance. Baker et al. (2007) indicated that the bright loops forming between the AR and the CH, the boundary retreat of the CH and the coronal dimming on one side of the AR were observational evidence of interchange reconnection occurring between a CH and an emerging AR. In addition to the evidence mentioned by them, Ma et al. (2014) also gave the other three evidence for interchange reconnection between a CH and an AR, including the brightening and EUV jets appearing at the CHBs and the disturbance of the AR. During the interaction between the CH and the filament in this event, we observed the brightenings at the boundaries and in the interior of the CH, the boundary retreat of the CH and the formation of dimmings at the two sides of the filament. Moreover, seeing from Figure 5(b), we found that the photospheric magnetic field at the feet of the expanding L and the CH's open field are favorable for Y-type interchange reconnection (as illustrated in Figure 4 of Wang \& Sheeley 2004) when L and the open field line are non-coplanar. Therefore, it is possible that the interaction between the expanding L and the CH's open field was interchange reconnection, which could lead to the appearance of brightenings, the boundary retreat of the CH and the formation of D1. It appears that, D2 and at least a portion of D1 should be rooted in the expansion of L and thus marked footprints of the expanding L. We think that the formation of D3 was due to density depletion caused by the eastward motion of the filament.

Along the PIL below the filament, horizontal strength values of the overlying magnetic field are measured at heights of 46 Mm and plotted in Figure 6(a). The transverse axis in the figure indicates the length of PIL which is scaled from its western to eastern endpoint to a range of 0-1. It is clear that the overlying magnetic field monotonically decreases from west to east, implying that the filament tended to activate toward a region with weak overlying field. Similar result occurring in the filament eruption event was also given by Bi et al. (2011). The gradient of magnetic field overlying filament has been believed to be an important factor in deciding the occurrence of torus instability and diagnosing full or failed filament eruption (Kliem \& T$\rm{\ddot{o}}$r$\rm{\ddot{o}}$k 2006; Liu 2008). It is usually voiced by decay index $n$ ($n$ = -$d$log($B$)/$d$log($h$), where $h$ is the height measured from the photosphere) of horizontal potential field. Based on the horizontal field strength values obtained from the PFSS model, we present the spatial rate of decay index $n$ along the PIL below the filament in Figure 6(b) to examine the gradient of the overlying magnetic field. Here, the range of $h$ is set to 46 to 100 Mm. Illustrated by the black solid line, the decay index shows mild drop from west to east, with the maximum value lower than 0.98. T$\rm{\ddot{o}}$r$\rm{\ddot{o}}$k \& Kliem (2005) gave a critical value of 1.53 for the occurrence of the torus instability. Kliem \& T$\rm{\ddot{o}}$r$\rm{\ddot{o}}$k (2006) indicated that the critical decay index ranges from 1.5 to 2.0 for torus instability. Liu (2008) argued that the decay index of a full filament eruption is greater than 1.74 and that of a failed eruption is lower than 1.71. Therefore, the decay index lower than 0.98 in our event could not satisfy the torus instability condition. The variation of the magnetic field overlying the filament was too weak to cause filament eruption.
	
%%%%%%%%%%%%%%%%%%%%%%%%%%%%%%%%%%%%%%%%%%%%%%%%%%%%%%%%%%%%%%%%%%%%%%%%%
\begin{figure}
\centering
\includegraphics[width=7cm]{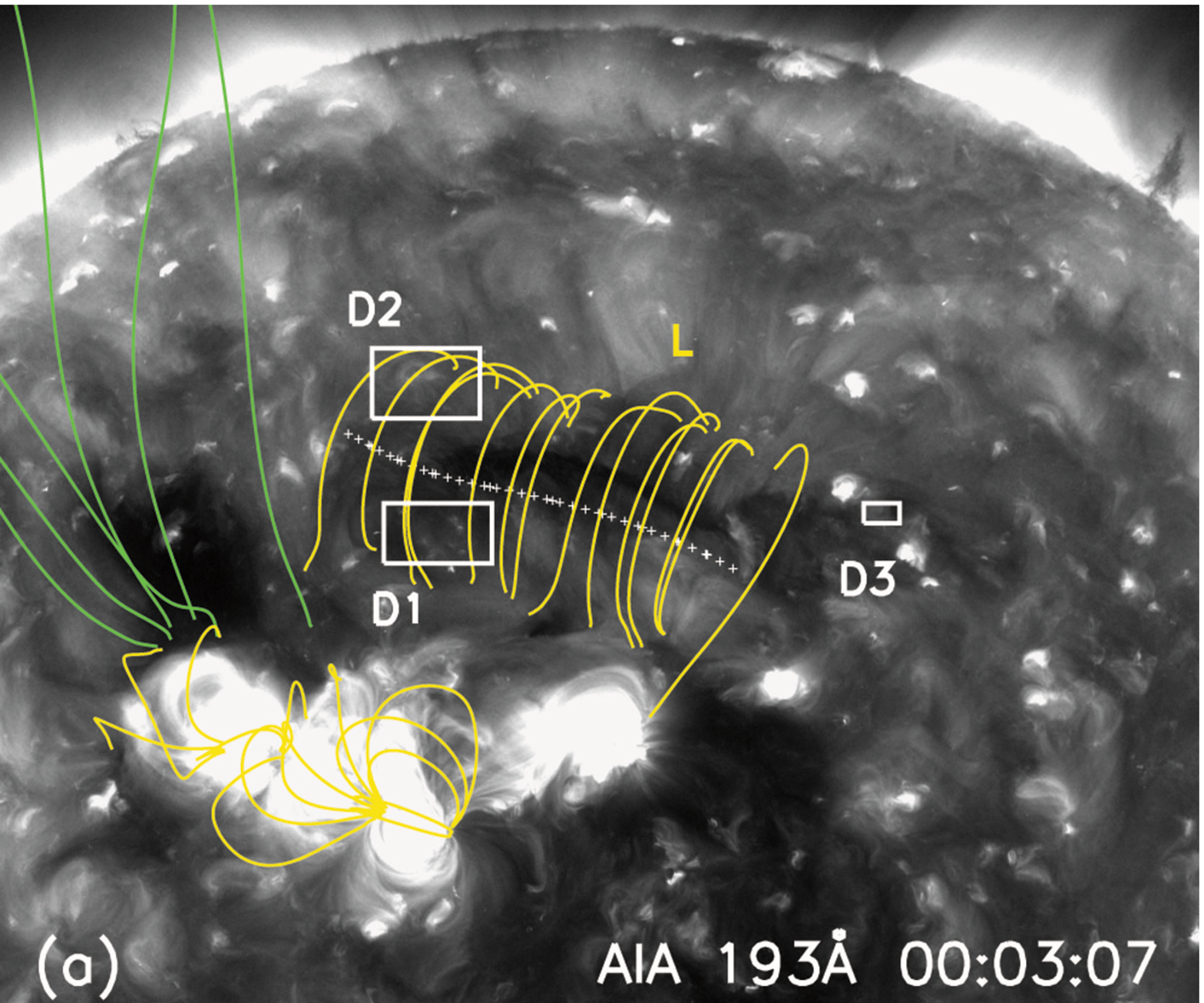}
\includegraphics[width=7cm]{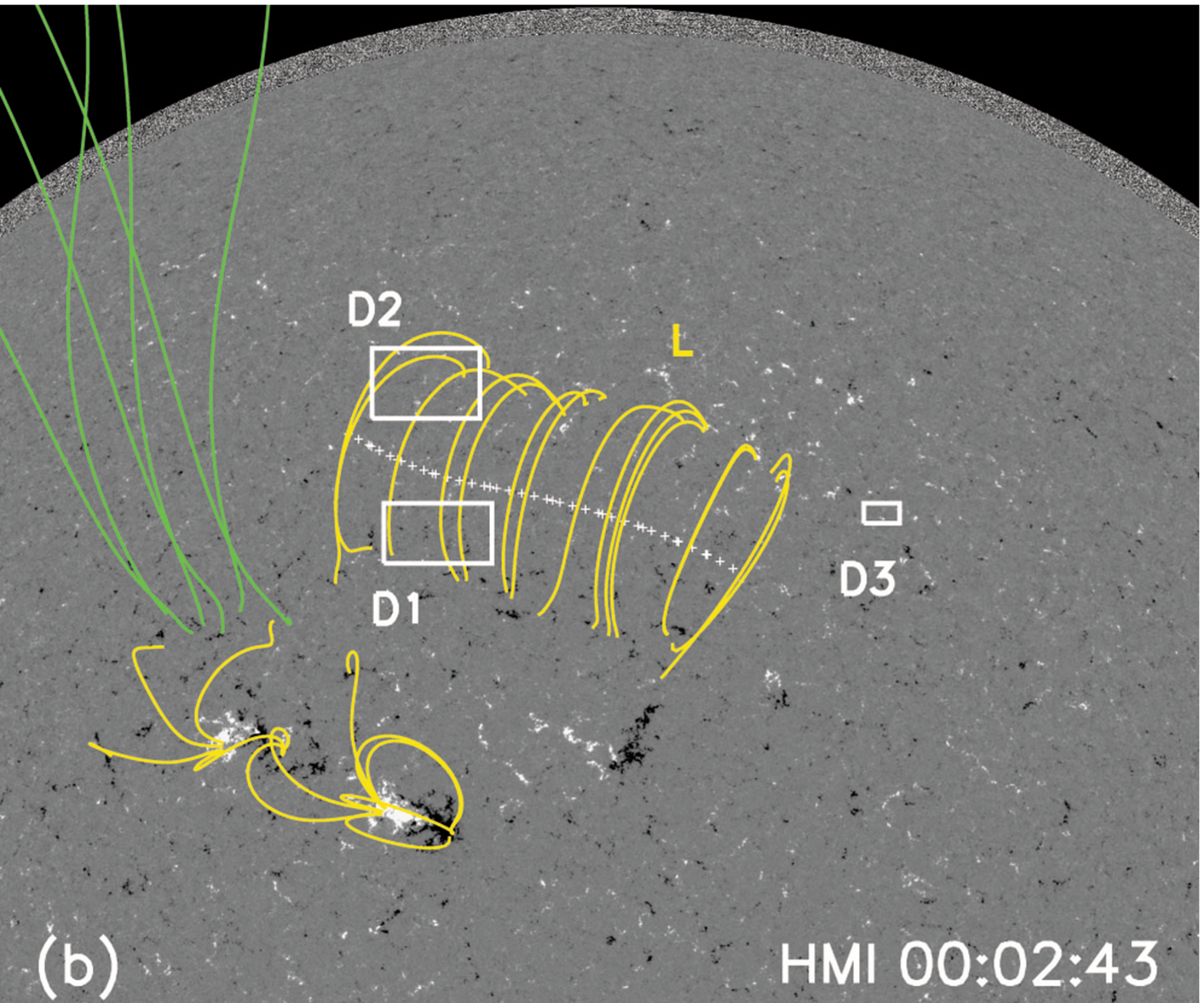}
\caption{AIA 193 {\AA} direct image (a) and HMI magnetogram (b) with the extrapolated field lines obtained from the PFSS model. The outlines of the three coronal dimmings represented by the white boxes and the PIL below the filament (white cross line) are overlaid. The yellow arcade, ``L,'' is linking D1 and D2. The green field lines refer to open lines of the CH.
The FOV is 1200$^{''}$ $\times$ 1000$^{''}$. The X range is from -600$^{''}$ to 600$^{''}$ and Y range is from 0$^{''}$ to 1000 $^{''}$.}
\label{f5}
\end{figure}

\begin{figure}
\centering
\includegraphics[width=10cm]{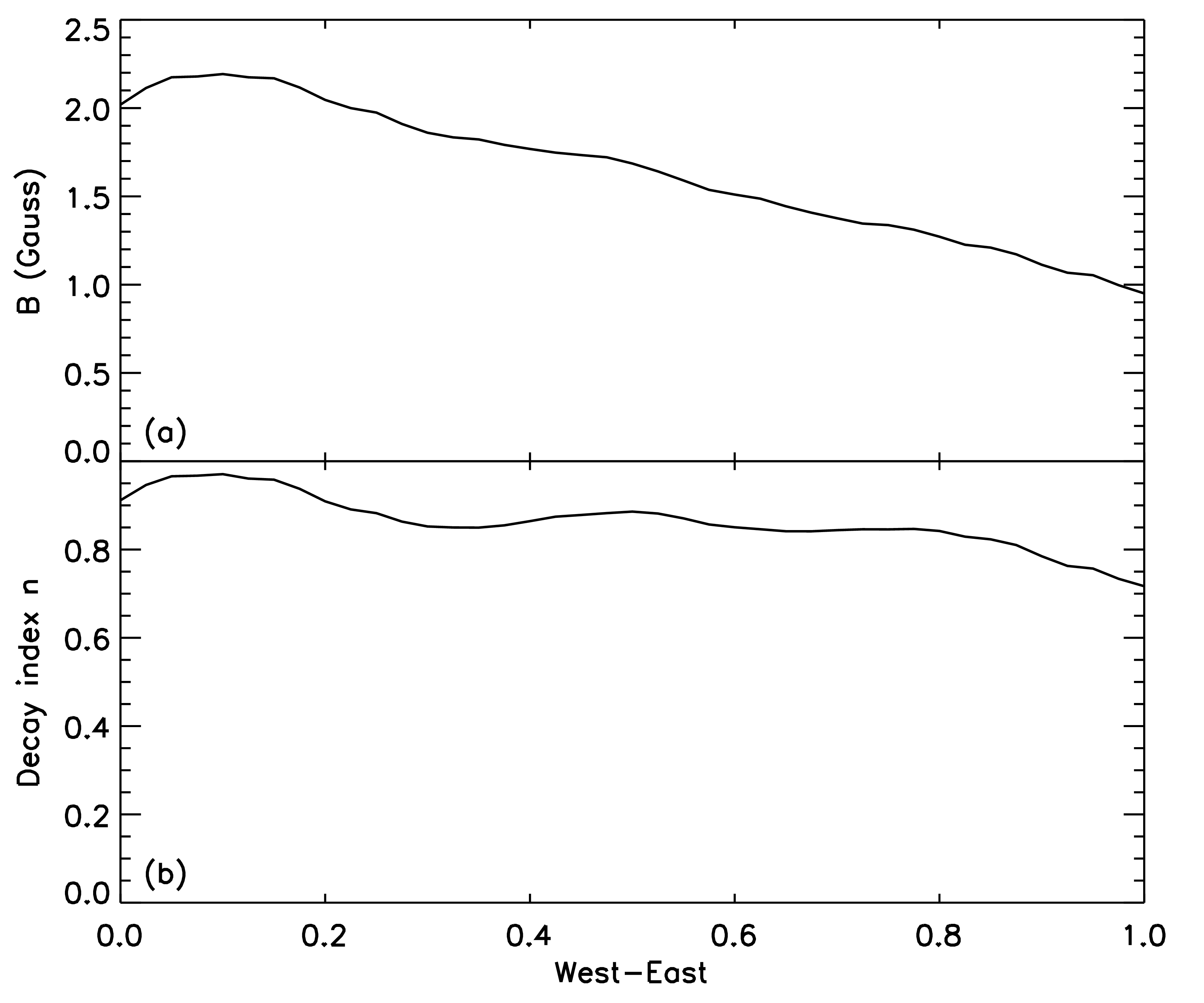}
\caption{The distribution of the overlying horizontal magnetic field strength at 46 Mm (solid curve in (a)) and the decay index estimated from 46 Mm to 100 Mm (solid line in (b)) along the PIL (shown in Figure (5)). The X-axis exhibits the length of the
PIL which is scaled from western to eastern endpoint to a range of 0-1.}
\label{f6}
\end{figure}
%%%%%%%%%%%%%%%%%%%%%%%%%%%%%%%%%%%%%%%%%%%%%%%%%%%%%%%%%%%%%%
\section{CONCLUSIONS AND DISCUSSIONS}

We present the interaction between the CH and the activating filament observed by the SDO and the GONG on 2010 November 13. The interaction was manifested as the intrusion of filament material to the CH, the appearance of the brightenings and the change of the CH. Comparing the result of the derived coronal magnetic configuration, the magnetic connectivity between D1 and D2 suggested that they were caused by the expansion of the overlying loop system forced by the filament activation. The interaction between the filament and the CH involved two processes: the direct intrusion of the filament to the CH, and the interaction between the expanding loop system and the CH's open field. Both of them could lead to the CH boundary retreat, however, the final disappearance of the CH resulted from the latter. The observations and the photospheric magnetic field setting reveal that, the interaction between the expanding loop system and the CH's open field might be interchange reconnection. In particular, the appearance of the brightenings, the CH boundary retreat and the partial formation of D1 might represent observational signatures of such reconnection. By estimating the horizontal strength and the decay index of the magnetic field overlying the filament, we found that the filament tended to activate toward a region with weak field and the variation of the magnetic field was too weak to induce filament eruption.
																			
Our observations are different from the case described by Jiang et al. (2007). The interaction in their work was due to the direct collision of the filament with the CH, and the filament was reflected back, with no or little filament material invading the CH. However, the interaction between the filament and the CH in our study was due to the direct intrusion of filament material to the CH and the expansion of the overlying loop system forced by the filament activation. We infer that the expanding loop system interacted with the CH's open field since the CH still changed after the filament intrusion and D1 and D2 persisted throughout the whole disappearance of the CH. This interaction might be Y-type interchange reconnection (Wang \& Sheeley 2004). The boundary retreat of the CH, the brightenings appearing at the boundaries and in the interior of the CH and the partial formation of D1 might be observational evidence of the reconnection.

Large-scale coronal magnetic configuration can play a significant role in different kinds of solar activities. As argued by Jiang et al. (2011), coronal dimmings can physically connect sympathetic filament eruptions. Yang et al. (2012) showed that a multiple-arcade bipolar streamer was in favor of the occurrence of sympathetic eruptions. In this event, the overlying loop system could expand to form coronal dimmings and interact with other magnetic structure. Our observations also reveal that the decay index lower than 0.98 was too small to cause filament eruption. The filament/CH interaction may have diverse forms. New forms and detailed mechanism are needed to be found out by means of more observations.

\begin{acknowledgements}
We thank the referee for many constructive suggestions that improved the quality of this paper. We are grateful to the SDO teams for providing the data. SDO is a mission for NASA's Living with a star (LWS) program. We also thank the GONG and GOES teams for data support. This work is supported by the National Science Foundation of China (NSFC) under grant numbers 11373066, 11373065, 11178016, Yunnan Science Foundation of China under number 2013FB086, the Talent Project of Western Light of Chinese Academy of Sciences, the National Basic Research Program of China 973 under grant number G2011CB811400.
	
\end{acknowledgements}

\label{lastpage}

\end{document}